# ОСОБЕННОСТИ СИНТЕЗА МОНОКРИСТАЛЛОВ $TbCr_3(BO_3)_4$ РАСТВОР-РАСПЛАВНЫМ МЕТОДОМ

©2023г. И.А. Гудим [1*], Н.В. Михашенок [1,2],

[1]*Институт физики им. Л.В. Киренского СО РАН, Красноярск, Россия*

[2]*Федеральный Иисследователлский Центр КНЦ СО РАН, Красноярск, Россия*

*irinagudim@mail.ru

Изучено фазообразование тербиевого хром-бората в растворах-расплавах на основе тримолибдата висмута и вольфрамата лития. Показано отсутствие тригональной фазы тербиевого хром-бората в системе на основе тримолибдата висмута при всех соотношениях компонентов. Найдено соотношение компонентов системы на основе вольфрамата лития, при которой при температурах выше 1100 $^o$C образуются тригональные кристаллы $TbCr_3(BO_3)_4$, а ниже этой температуры только моноклинная фаза. Изучены рентгеновские свойства выращенных кристаллов.

## ВВЕДЕНИЕ

В настоящее время большой интерес проявляется к кристаллам с мультиферроичными свойствами. Изучается взаимосвязь проявляемых свойств со структурой и составом соединений. К числу таких соединений, проявляющих различные мультиферроичные свойства, относится и семейство тригональных редкоземельных оксиборатов со структурой хантита, пространственная группа R32 (P312$_1$) [1-3]. Это семейство интересно своей высокой вариабильностью как на позиции лантаноидов, так и на позиции переходных элементов. Изначально эти кристаллы привлекли внимание как среды для лазеров с самоудвоением частоты. Уже в 80-е годы прошлого века активно велись исследования в этой области[4-5]. Не менее интересны были и их нелинейно-оптические свойства [6-8] И лишь позднее обратили внимание на их магнитные и пьезоэлектрические свойства.

Уже более 20 лет активно изучаются кристаллы ферроборатов со структурой хантита [9-11]. Несколько меньше изучаются кристаллы алюмо-, галло и скандоборатов [12-14]. Еще одним подсемейством являются хромовые соединения со структурой хантита.

Мы обратились к кристаллам $TbCr_3(BO_3)_4$ поскольку только у соединений на основе ионов тербия (ферро-, алюмобораты) магнитоэлектрические и магнитооптические свойства проявляются вплоть до комнатных температур, в то время как у соединений с другими редкоземельными ионами они исчезают при гораздо более низких температурах.

Ранее выращивание кристаллов хромовых оксиборатов традиционно проводилось из растворов-расплавов на основе молибдата калия $K_2MoO_4$ и оксида бора [15-17]. Однако в работе [16] авторы отмечают, что часть выращенных кристаллов содержала до 10% другой фазы. Кроме того, спектроскопические исследования проведенные на монокристаллах иттриевого алюмобората, допированного иттербием, выращенных из различных раствор-расплавных систем, показали достаточно большое количество примесей, как калия, так и молибдена при росте из калий-молибдатной системы[18]. В той же работе показано, что при росте из висмут-молибдатного раствора вхождение примесей существенно меньше. Поэтому первой системой, из которой мы пытались вырастить монокристаллы тербиевого хромбората была именно такая система.

## КРИСТАЛЛИЗАЦИЯ ИЗ ВИСМУТ-МОЛИБДАТНОГО РАСТВОРА-РАСПЛАВА.

Из опыта выращивания алюмо- и ферроборатов со структурой хантита был выбран растворитель с молярным соотношением тримолибдата висмута к оксиду бора $Bi_2Mo_3O_{12} : B_2O_3 = 1:2$. И стехиометрической концентрацией кристаллообразующих окислов 10% вес. Температура насыщения такого раствора-расплава оказалась выше 1200 °C. Но единственной кристаллизующейся фазой вплоть до T = 800 °C оказался оксид хрома. Постепенно увеличивали молярное соотношение оксида бора к тримолибдату висмута до 3:1, и именно при таком соотношении поменялась кристаллизующаяся фаза, ею стал борат хрома $CrBO_3$. Следующим шагом стало введение сверх стехиометрии оксида тербия $Tb_2O_3$. После того как молярное соотношение оксида тербия к тримолибдату висмута достигло величины 0,3:1, снова сменилась кристаллизующаяся фаза, появился $TbCr_3(BO_3)_4$, но к сожалению только моноклинный с пространственной группой C2/c. Дальнейшие изменения раствора-расплава не привели к появлению тригональной фазы. Окончательный вид раствора-расплава в квазибинарной форме можно записать как:

92%вес {$Bi_2Mo_3O_{12} + 3B_2O_3 + 0,3\ Tb_2O_3$} + 8%вес $TbCr_3(BO_3)_4$ (1)

Из данного раствора-расплава методом спонтанного зарождения с последующим медленным (1-3 °C/сутки) снижением температуры были выращены моноклинные кристаллы тербиевого хромбората для исследования их физических свойств.

КРИСТАЛЛИЗАЦИЯ И РОСТ ИЗ ЛИТИЙ_ВОЛЬФРАМАТНОГО РАСТВОРА-РАСПЛАВА.

Другим изученным нами растворителем стал вольфрамат лития $Li_2WO_4$, использовавшийся ранее для выращивания монокристаллов тригональных редкоземельных алюмо-, скандо- и ферроборатов. Выращивание из растворов-расплавов на основе вольфрамата лития позволило получить кристаллы с меньшим количеством примесей компонентов раствора-расплава [19].

Изначально исследовалась система с молярным соотношением компонентов растворителя $Li_2WO_4:B_2O_3:Tb_2O_3 = 1:3,3:0,3$ и стехиометрической концентрацией кристаллообразующих окислов 15 вес.%. Температура насыщения в таком растворе-расплаве оказалась равной $1090^{o/}C$. Образовались мелкие кристаллы хантита вблизи поверхности раствора-расплава и еще более мелкие кристаллы $Cr_2O_3$ вблизи дна тигля.

Области стабильности этих кристаллов, а также соотношения компонентов растворов-расплавов были определены методом прямого фазового зондирования. Температура насыщения ($T_{нас}$) определялась с точностью $\pm 5^0C$ с помощью пробных кристаллов, которые предварительно получали из того же раствора-расплава в условиях спонтанного зарождения на вращающийся платиновый стержневой держатель. Ширина метастабильной зоны $\Delta T_{мет} \approx 12^0C$ определялась как максимальное переохлаждение, при котором не было зарождения за 20-часовой период времени.

Растворы-расплавы массой 100г готовились при $T=1100^0C$ в платиновом цилиндрическом тигле (D=60мм, H=65мм) сплавлением окислов [$Li_2WO_4$, $B_2O_3$, $Tb_2O_3$, $Cr_2O_3$] в соотношении, определяемом выше приведённой формулой. Тигель устанавливался в кристаллизационную печь, где температура уменьшалась от дна тигля с вертикальным градиентом 1-2$^0$C/см. Раствор-расплав гомогенизировался при T=1100$^0$C за 24 часа. Для поддержания однородности раствор-расплав перемешивался.

Для получения затравок кристаллов TbCr$_3$(BO$_3$)$_4$ использовался метод спонтанного зарождения. Для этого, после определения параметров кристаллизации, в раствор-расплав при температуре гомогенизации погружаем стержень и включаем вращение со скоростью 40 об/мин. Через 2 часа температура раствора-расплава понижается на 15$^o$C ниже температуры насыщения. Затравки разращиваются в течение суток. Затем стержень извлекается из печи. Остатки раствора-расплава удаляются кипячением в 20% водном растворе азотной кислоты. Выросшие кристаллы-затравки снимаются со стержня и используются в дальнейшем для выращивания более крупных кристаллов.

После этого 4 визуально качественные затравки были закреплены на платиновом стержневом держателе. Держатель погружался в раствор-расплав при температуре T=T$_{нас}$+7$^0$C и включалось реверсивное с периодом 1мин вращение со скоростью ω=30 об/мин. Через 15 мин температуру понизили до T=T$_{нас}$-7$^0$C. Далее температура раствора-расплава снижалась с нарастающим темпом 1-2$^0$C в сутки, так чтобы скорость роста кристаллов не превышала 0,5 мм в сутки. Рост продолжался 8 дней. После завершения процесса роста держатель приподнимался над поверхностью раствора-расплава и охлаждалось до комнатной температуры с выключенным питанием печи. В результате выросли 3 кристалла размером 3x5 мм.

ХАРАКТЕРИЗАЦИЯ ПОЛУЧЕННЫХ ОБРАЗЦОВ

На полученных образцах был проведен рентгеноструктурный анализ кристаллической структуры на монокристальном дифрактометере SMART APEX II (Bruker AXS, установка ЦКП КНЦ СО РАН). Измерения проводились на монокристаллах, полученных с использованием разных растворителей, упомянутых в данной работе. Так, при использовании в качестве растворителя тримолибдата висмута $BiMo_3O_{12}$, были получены монокристаллы с пространственной группой C2/c. Для уточнения параметров кристаллической структуры исследования проводились на поликристаллическом образце, полученном перетиранием монокристалла в порошок. Использование вольфрамата лития $Li_2WO_4$ позволило найти стабильные области роста тригональной фазы монокристаллов хантита с пространственной группой R32. Параметры элементарных ячеек приведены в Таблице 1.

ЗАКЛЮЧЕНИЕ

Изучено фазообразование тербиевого хром-бората в растворах-расплавах на основе тримолибдата висмута и вольфрамата лития. Показано отсутствие тригональной фазы тербиевого хром-бората в системе на основе тримолибдата висмута при всех соотношениях компонентов. Найдено соотношение компонентов системы на основе вольфрамата лития, при которой при температурах выше 1100 $^o$C образуются тригональные кристаллы $TbCr_3(BO_3)_4$, а ниже этой температуры только моноклинная фаза. Изучены рентгеновские свойства выращенных кристаллов.

## БЛАГОДАРНОСТЬ

Таблица 1. Параметры элементарных ячеек выращенных кристаллов.

| Симметрия | Параметры элементарной ячейки, Å or ° |
|---|---|
| R32 | a=b=9.47; c=7.49; |
| C2/c | a=7.41; b=9.65; c=11.31; β=103.42. |

Литература

1. Liang K.-C., Chaudhury R. P., Lorenz B., et al. // PHYSICAL REVIEW B 83. 2011. 180417(R).

2. Usui, T, Tanaka, Y, Nakajima, H, Taguchi, et al. // Nature Materials 13, 2014, 611.

3. Goldner Ph., Guillot-Noël O., and Petit J., et. al. //Phys. Rev. B76, 2007, 165102.

4. Jinsheng Liao, Yanfu Lin, Yujin Chen, et al. // Journal of Crystal Growth 267, 2004, 134–139.

5. Dorozhkin L.M., Kurstev I.I., Leonyuk N.I., et. al. // Sov. Thech. Phys. Lett. 7, 1981, 555-556.

6. Chaoyang Tu, Zhaojie Zhu, Jianfu Li, et.al. // Optical Materials 27, 2004, 167–171.

7. Ran He, Lin Z. S., Lee M.-H., and Chen C. T // JOURNAL OF APPLIED PHYSICS 109, 2011, 103510.

8. Tu C., Journal of Crystal Growth 208, 2000, 487.

9. Popova E.A., Volkov D.V., Vasiliev A.N., et.al. //Phys. Rev. B 75, 2007, 224413.

10. Volkov D.V., Popova E.A., Kolmakova N.P., et.al. //JMMM, 316, 2007, e717-e720.

11. Звездин А.К., Воробьев Г.П., Кадомцева А.М et.al. // Письма в ЖЭТФ, т. 83, 2006, в. 11, с. 600-605. ; JETP Letters, 2015, Vol. 101, No. 5, pp. 318–324

12. Александровский А.С., Гудим И.А., Крылов А.С., et al. //ФТТ, т. 49, в. 9, 2007, с. 1618-1621.
13. Volkov N. V., Gudim I. A., Eremin E. V., et al. // Письма ЖЭТФ 99, 2014, No. 2, с. 72–80; JETP Letters, 2015, Vol. 101, No. 5, pp. 318–324
14. Eremin E.V., Pavlovskiy M.S., Gudim I.A., et al. // Journal of Alloys and Compounds 828, 2020, 154355
15. Wang G.F. Non-linear Optical Crystals I. Structure and Bonding, vol. 144, Springer, Berlin, 2012, pp. 105-120.
16. Leonyuk N.I., Maltsev V.V., Volkova E.A et al. // Opt. Mater. 30 (1), 2007, 161-163.
17. Yongyuan Xu, Xinghong Gong, Yujin Chen, et a. // Journal of Crystal Growth 252, 2003, 241–245.
18. Boldyrev K.N., Popova M.N., Bettinelli M et al. // Optical Materials 34, 2012, 1885–1889.
19. Eremin E.V., Pavlovskiy M.S., Gudim I.A., et al. // Journal Of Alloys And Compounds. DOI: https://doi.org/10.1016/j.jallcom.2020.154355

# FEATURES OF SYNTHESIS OF TbCr3(BO3)4 SINGLE CRYSTALS BY SOLUTION-MELT METHOD

©2023 I.A. Gudim [1*], N.V. Mikhashenok [1,2]

[1]Institute of Physics named after. L.V. Kirensky SB RAS, Krasnoyarsk, Russia

[2]Federal Research Center KSC SB RAS, Krasnoyarsk, Russia

*irinagudim@mail.ru

The phase formation of terbium chromium borate in melt solutions based on bismuth trimolybdate and lithium tungstate was studied. It was shown that there is no trigonal phase of terbium chromium borate in a system based on bismuth trimolybdate at all component ratios. The ratio of components of a system based on lithium tungstate has been found, at which trigonal $TbCr_3(BO_3)_4$ crystals are formed at temperatures above 1100 °C, and below this temperature only a monoclinic phase is formed. The X-ray properties of the grown crystals were studied.

Keywords: phase formation in melt solutions, terbium chromoborate, spontaneous crystallization, growth on seeds